# Remarkable band gap renormalization via dimensionality of the layered material $C_3B$

Yabei Wu,[†,a,c] Weiyi Xia,[†b] Yubo Zhang,[a] Wenguang Zhu,[c] Wenqing Zhang,[*a] and Peihong Zhang[*b]

[a] Department of Physics, Shenzhen Institute for Quantum Science and Engineering, and Academy for Advanced Interdisciplinary Studies, Southern University of Science and Technology, Shenzhen, Guangdong 518055, China

[b] Department of Physics, University at Buffalo, State University of New York, Buffalo, New York 14260, USA

[c] ICQD, Hefei National Laboratory for Physical Science at the Microscale, Key Laboratory of Strongly-Coupled Quantum Matter Physics, Chinese Academy of Sciences, Department of Physics, and Synergetic Innovation Center of Quantum Information and Quantum Physics, University of Science and Technology of China, Hefei, Anhui 230026, China

[†] Y. W. and W. X. contributed equally to this work.

[*] E-mails: zhangwq@sustech.edu.cn; pzhang3@buffalo.edu

## Abstract

Layer-dependent electronic and structural properties of emerging graphitic carbon boron compound $C_3B$ are investigated using both density functional theory and the GW approximation. We discover that, in contrast to a moderate quasiparticle band gap of 2.55 eV for monolayer $C_3B$, the calculated quasiparticle band gap of perfectly stacked bulk phase $C_3B$ is as small as 0.17 eV. Therefore, our results suggest that layered material $C_3B$ exhibits a remarkably large band gap renormalization of over 2.3 eV due to the interlayer coupling and screening effects, providing a single material with an extraordinary band gap tunability. The quasiparticle band gap of monolayer $C_3B$ is also over 1.0 eV larger than that of $C_3N$, a closely related two-dimensional semiconductor. Detailed inspections of the near-edge electronic states reveal that the conduction and valence band edges of $C_3B$ are formed by out-of-plane and in-plane electronic states, respectively, suggesting an interesting possibility of tuning the band edges of such layered material separately by modulating the in-plane and out-of-plane interactions.

## I. Introduction

Despite unprecedented research efforts, practical applications of graphene in electronic devices remain distant. Fulfilling the promise of graphene as a building block of next generation electronic devices may hinge on our ability to introduce a stable and sizable band gap in graphene in a systematic and controllable manner. A number of approaches have been proposed to tune the electronic properties of graphene with varying degree of success [1-13]. Recently, carbon-nitrogen based graphene-like two-dimensional (2D) semiconductors such as $C_2N$ [14] and $C_3N$ [15-17] have attracted considerable research interest. The electronic structure of $C_3N$ can be understood by shifting the Fermi level from that of graphene and the opening of a band gap due to the nitrogen potential. The quasiparticle band gap of monolayer $C_3N$ has been predicted to be about 1.5 eV [16], an ideal value for electronics applications. Considering the electron-hole symmetry of the low energy electronic structure of graphene, it is straightforward to speculate that monolayer hexagonal $C_3B$ should also be a 2D semiconductor with a moderate band gap.

Interestingly, graphitic $C_3B$ has been successfully synthesized in the bulk form [18]. The as-synthesized samples do not display a particularly dominant stacking pattern. Surprisingly, bulk $C_3B$ samples appear to be metallic [18]. The proposed structure [18] of monolayer $C_3B$ was confirmed later using global optimization methods [19]. Electronic structure of bulk $C_3B$ has also been investigated [20] within density functional theory (DFT) assuming two stacking patterns. Both bulk phases were shown to be metallic within the local density approximation (LDA), which seems to be consistent with experiment. Monolayer $C_3B$, on the other hand, has been predicted to be a semiconductor with a small indirect band gap of 0.66 eV [20] within LDA. Considering that LDA almost always underestimates the band gap of *sp* semiconductors, especially for 2D systems, we expect that the true band gap of monolayer $C_3B$ to be much larger. In addition, it would also be interesting to understand how the presumably weak interlayer interaction could render a moderate-gap 2D semiconductor metallic in the bulk form.

Accurate understanding of the electronic structure of $C_3B$ requires advanced computational methods going beyond DFT with local or semilocal functionals. In this work, we used the GW method [21] to systematically investigate the electronic structures of $C_3B$ from monolayer to bulk phase. The calculated GW band gap of monolayer $C_3B$ is about 2.55 eV, which is about 1.9 eV larger than the DFT result. This band gap is also 1.0 eV larger than the band gap of monolayer $C_3N$ [16]. We further investigate the effects of interlayer interactions on the electronic structure of $C_3B$ by constructing four stacking models for the bilayer and bulk phases. Similarities and differences between $C_3B$ and $C_3N$ are discussed. All four bilayer models have moderate band gaps ranging from 1.66 to 1.89 eV depending on the atomic registry, giving rise to a large band gap reduction of 0.7 ~ 0.9 eV compared with that of monolayer. This reduction in band gap is significant, given that the interlayer separation is greater than 3.2 Å. All four bulk models are predicted to be semimetal within DFT. However, subsequent GW calculations reveal that perfectly stacked bulk $C_3B$ may actually be narrow gap semiconductor with an indirect band gap of about 0.2 eV. Our results call for future experimental verifications.

## II. Computational Methods

Structure optimizations are carried out using the Vienna Ab-initio Simulation Package (VASP) [22,23]. The ion-electron interaction is treated using the projected augmented-wave (PAW) technique [24]. Three functionals, i.e., the Perdew-Burke-Ernzerhof (PBE) functional [25], and two van der Waals (vdW) functionals, optB86b-vdW [26] and SCAN+rVV10 [27] are used in this work for structural optimizations. The structures are fully optimized until the maximum energy and force are less than $10^{-6}$ eV and 0.01 eV/Å, respectively. The Brillouin zone (BZ) integration is carried out using a 9×9×1 uniform $k$-grid for the monolayer and bilayer systems and 9×9×8 for the bulk models. A vacuum space of about 20 Å is introduced for the monolayer and bilayer models.

We use the structures optimized with the PBE functional (for monolayer $C_3B$) or the optB86b-vdW [26] (for bilayer and bulk systems) for subsequent GW calculations. The quasiparticle band structures are calculated within the $G^0W^0$ (i.e., one-shot GW) approach [21] using a local version of the BERKELEYGW package [28] based on version 1.0.4 in which recently developed acceleration methods [16,29-31] are implemented. Our methods allow carrying out highly converged GW calculations for 2D materials with a fraction of computational costs compared with the conventional approach as we have discussed in our previous work [16,29]. The energy-integration method [30,31] we developed greatly alleviates the burden of band summation in the GW calculations, effectively reducing the computational cost by over one order of magnitude. We have also developed a combined sub-sampling and analytical integration method to combat the slow convergence [32-35] of the GW self-energy with respect to the BZ sampling density for 2D materials. Using this method, we are able to achieve well converged GW results using a 6×6×1 $k$-grid for 2D materials [16,29]. We use the Hybertsen-Louie generalized plasmon-pole model (HL-GPP) [21] to extend the static dielectric function to finite frequencies. The HL-GPP model has been used in GW calculations for a wide-range of systems from solids to molecules with great success. In particular, there have been several published works [34,36-38] on GW calculations of 2D systems using the HL-GPP models. Other details of our GW calculations are discussed later. We would like to mention that all GW calculations are carried out assuming static crystal structures. Therefore, our results do not include possible electron-phonon renormalization effects [39-41] which, if included, will result in a small reduction to the calculated band gap. We expect our main results remain valid for monolayer and bilayer systems. For bulk $C_3B$, however, since the calculated quasiparticle band gaps are already very small, electron-phonon renormalization and temperature effects may render the material to be semi-metallic at room temperature.

## III. Results and Discussion

### A. Monolayer $C_3B$.

The optimized crystal structure of monolayer $C_3B$ is shown in Figure 1a. The in-plane lattice constant of $C_3B$ is 5.166 Å optimized using the PBE [25] functional. This is about 6.2% larger than that of monolayer $C_3N$. The C-B bonds (1.562 Å) are substantially longer than the C-N bonds (1.401 Å) in $C_3N$. In fact, even the C-C bonds in $C_3B$ are slightly longer than those in $C_3N$ (1.420 vs 1.405 Å). Figure 1b shows the DFT-PBE and GW band structures of monolayer $C_3B$; Figure 1c shows the band structure of $C_3N$ for comparison. The DFT-PBE band gap of $C_3B$ (0.64 eV) is slightly larger than that of $C_3N$ (0.39 eV). The GW band gap, however, is significantly larger (2.55 vs 1.50 eV). The greatly enhanced GW band gap of $C_3B$ comes from a surprisingly large QP correction for the VBM states at the $\Gamma$ point. We will come back to this point later. Another interesting observation is that, although $C_3B$ and $C_3N$ are both indirect band gap semiconductors, the valence band maximum (VBM) and conduction band minimum (CBM) positions are switched in two systems: The CBM and VBM of $C_3B$ ($C_3N$) are located at the M ($\Gamma$) point and $\Gamma$ (M) point, respectively. We mention that due to the significant changes to the near edge states, some of the novel physics of graphene (e.g., Dirac cone dispersion and pseudospin) is not present in $C_3B$, unless under heavy electron doping condition that brings the Fermi level significantly above the band gap to near the Dirac point.

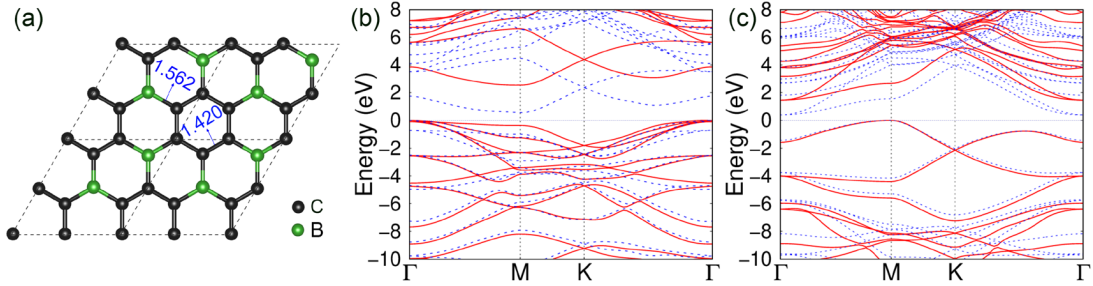

FIG. 1. Crystal structure (a) and band structure (b) of monolayer $C_3B$. The band structure of $C_3N$ (c) is also shown for comparison. GW results are shown with red solid lines whereas DFT results are shown with blue dotted lines.

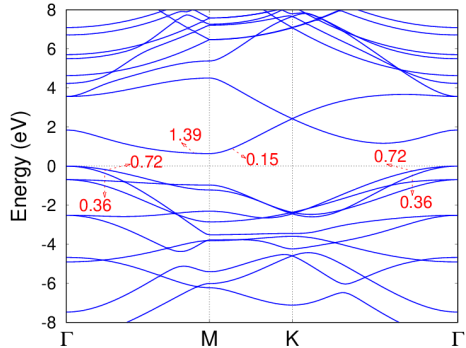

FIG. 2. Calculated electron and hole effective masses of $C_3B$.

Figure 2 shows the calculated effective masses at the valence and conduction extrema. Interestingly, the effective mass of the heavy hole state is about two times that of the light hole state. Although the hole effective masses are isotropic as guaranteed by the $C_3$ symmetry of the system, the effective mass of the CBM state (at the M point) is highly anisotropic. The electron effective mass along the M→K direction (i.e., transverse mass

$m_{tr}$, corresponding to the zigzag direction in the real space) is much small than that along the M→Γ direction (longitudinal mass $m_l$).

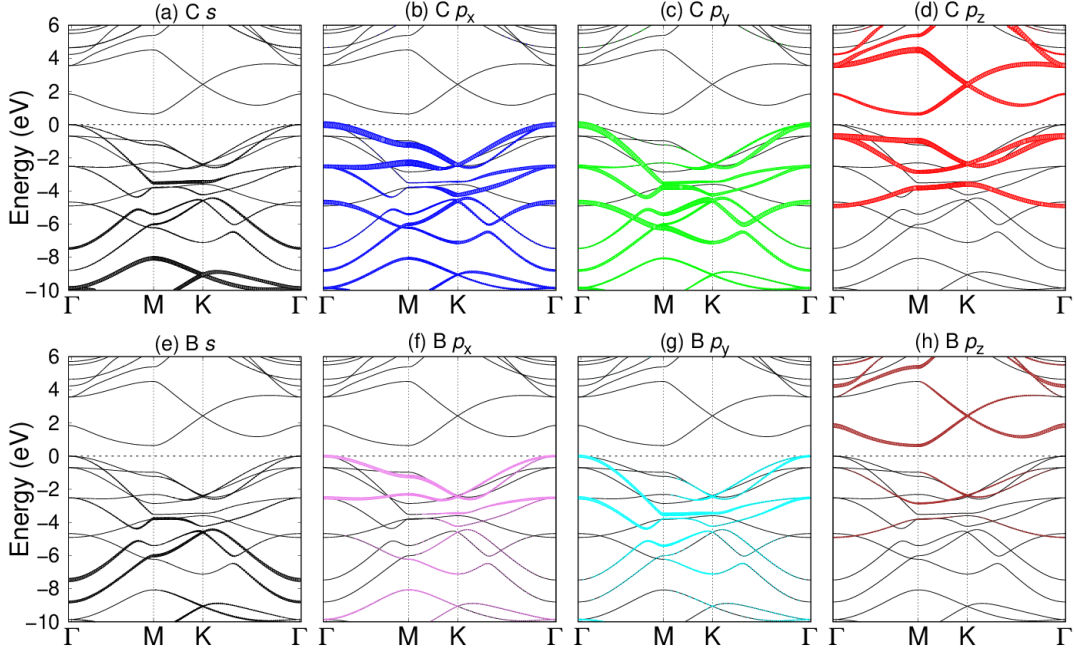

FIG. 3. Decomposition of the Bloch wave functions into contributions from atomic orbitals. The CBM states are mostly derived from the carbon and boron $p_z$ orbitals whereas the VBM states are dominated by contributions from the in-plane $p_x$ and $p_y$ orbitals.

To gain more understanding of the properties of the band edge states, we show in Figure 3 the decomposition of the Bloch wave functions into atomic orbital contributions. Whereas the low energy conduction bands are derived almost exclusively from the out-of-plane $p_z$ (i.e., $p\pi$) states, near-edge valence states have contributions from both in-plane and out-of-plane $p$ orbitals. Interestingly, only the in-plane $p_x$ and $p_y$ (i.e., σ) states contribute to the VBM at the Γ point. Therefore, the band gap is formed between out-of-plane (CBM) and in-plane (VBM) states. This raises an interesting possibility of tuning the VBM and CBM states separately by in-plane and out-of-plane interactions since the σ states are more sensitive to in-plane stress whereas the π states are more susceptible to interlayer interactions. We expect that the conduction states would be affected strongly if two $C_3B$ layers are brought together to form bilayer systems. In fact, this explains in part the large band gap reduction going from monolayer to bilayer systems as we will discuss in the next section. The conduction states can be further tuned if the inter-plane separation is modified (for example, by applying uniaxial pressure). On the other hand, valence band edge states can be manipulated by applying in-plane stress. We will come back to this point in later. Future experiments (or applications) may exploit these results to tune conduction or valence states separately by applying in-plane or out-of-plane strains. In addition, when $C_3B$ layers are stacked to form bulk phase, the interlayer interaction may lead to overlaps between the π and σ states. This may give rise to simultaneous presence of σ and π bands at the Fermi level and may explain the enhanced conductivity in bulk $C_3B$ [42] compared with graphite. In contrast, in the case of $C_3N$, both CBM and VBM states are derived exclusively from $p_z$ orbitals [16].

The fact that the VBM states of $C_3B$ have the in-plane (more localized) $p\sigma$ character also helps to explain the enhanced quasiparticle corrections to the band gap of monolayer $C_3B$ compared with that of $C_3N$ as mentioned earlier. As it is shown in Figure 3 (b), (c), and (d), the two-fold degenerate VBM states at the Γ point have the in-plane $p\sigma$ character [Figure 3 (b) and (c)], whereas the two valence states below the VBM have the $p\pi$ character [Figure 3 (d)]. The separation between these two doublets is about 0.7 eV at the DFT-PBE level. Upon applying the GW corrections, these two doublets are essential degenerate as can be seen in Figure 1(b) due to the greater quasiparticle correction for the in-plane $p\sigma$ states which pushes down the VBM states, leading to a much wider band gap compared with $C_3N$. The distinct quasiparticle correction for states with different atomic characters (i.e., in-plane vs out-of-plane wave functions) can also be seen from Figure 4. States with primarily in-plane wave functions clearly have greater quasiparticle corrections than out-of-plane ones. The contrasting near-edge electronic states between $C_3N$ and $C_3B$ will also lead to different interlayer interaction behaviors as we will discuss in the next section.

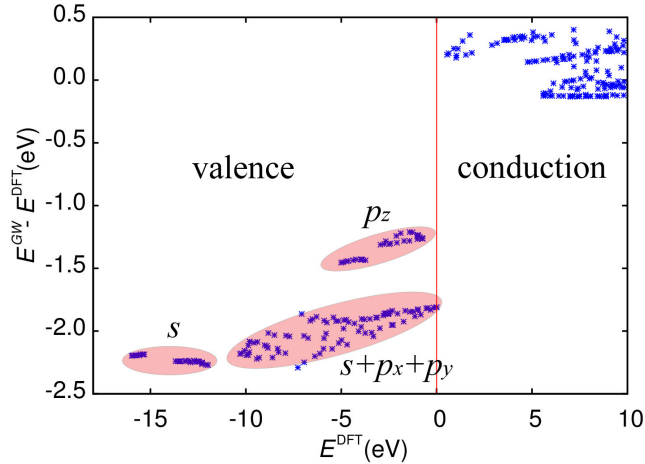

FIG. 4. Quasiparticle correction v.s. DFT energy plot showing distinct quasiparticle correction for states with different atomic character.

**B. Bilayer and bulk $C_3B$.**

We now investigate bilayer $C_3B$, considering four stacking models (named AA1, AA2, AB1, and AB2) as shown in Figure 5. In the AA1 stacking, atoms in the top layer are directly above the same atoms in the bottom layer. The AA2 stacking can be obtained by shifting the top layer in the AA1 stacking by half lattice constant along the diagonal direction so that interlayer B-B pairs are avoided. The AB1 (AB2) stacking is obtained by shifting the top layer in the AA1 (AA2) structure along the diagonal direction by 1/3 of the lattice constant.

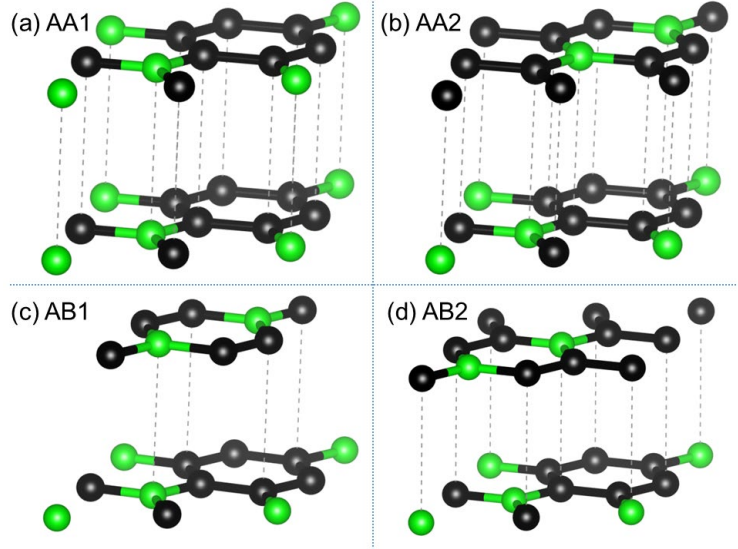

FIG. 5. Four bilayer models of $C_3B$ studied in this work.

Table I shows the optimized lattice parameters and total energy difference for the four stacking models using the PBE functional and van der Waals (vdW) functionals optB86b-vdW [26] and SCAN+rVV10 [27]. The in-plane lattice constants predicted using the optB86b-vdW and PBE functionals are very similar, which are about 0.5% greater than those predicted using the SCAN+rVV10 [27] functional. Both vdW-optB86b and SCAN+rVV10 predict an interlayer distance ranging from 3.3 to 3.7 Å, which is reasonable for these layered structures. We have also test the DFT+D3 [43] method and obtained results that are similar to those calculated using vdW-DF functionals. The PBE functional, on the other hand, obviously overestimates the interlayer separations. For simplicity, we will use the structures optimized using the optB86b-vdW functional for subsequent electronic structure calculations. Within the results obtained using vdW functionals, we notice that the interlayer distance of the AA2 model is significantly smaller (by over 0.3 Å) than that of the AA1 model. In addition, the AA2 stacking clearly has the lowest energy among the four studied models.

TABLE I. Optimized crystal structures ($a$ is the in-plane lattice constant and $d$ is the interlayer distance) and their relative energies (meV/atom) for the four bilayer models shown in Figure 5.

| Functional | PBE | | | vdW-optB86b | | | SCAN+rVV10 | | |
|---|---|---|---|---|---|---|---|---|---|
| Stacking | $a$ (Å) | $d$ (Å) | $\Delta E$ (meV) | $a$ (Å) | $d$ (Å) | $\Delta E$ (meV) | $a$ (Å) | $d$ (Å) | $\Delta E$ (meV) |
| AA1 | 5.168 | 4.667 | 0.26 | 5.165 | 3.667 | 10.73 | 5.137 | 3.703 | 9.08 |
| AA2 | 5.167 | 4.124 | 0.00 | 5.161 | 3.284 | 0.00 | 5.135 | 3.373 | 0.00 |
| AB1 | 5.165 | 4.461 | 11.31 | 5.162 | 3.496 | 17.97 | 5.135 | 3.556 | 15.98 |
| AB2 | 5.166 | 4.109 | 12.19 | 5.166 | 3.279 | 13.35 | 5.138 | 3.364 | 11.98 |

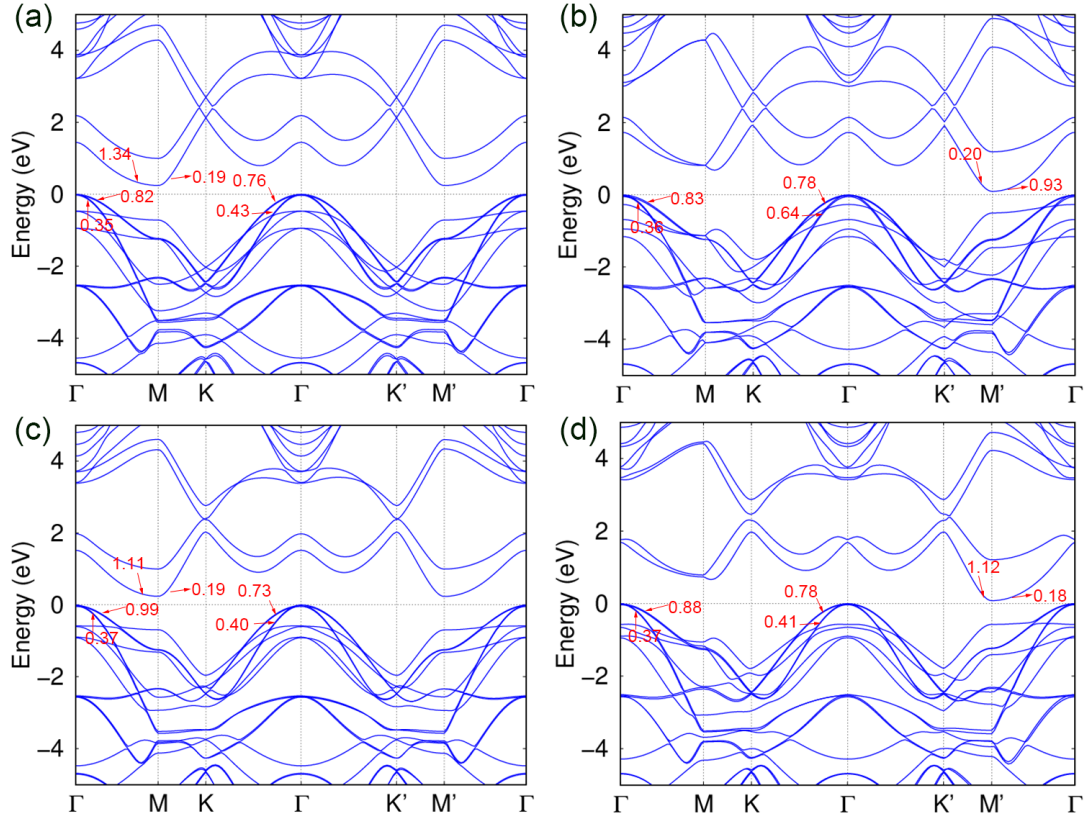

FIG. 6. DFT-PBE band structures of bilayer $C_3B$ showing also the electron and hole effective masses: (a) AA1, (b) AA2, (c) AB1, and (d) AB2.

Figure 6 shows the DFT-PBE band structures for the four bilayers systems with electron and hole effective masses show in the figures. The calculated DFT-PBE band gap varies from 0.08 (AB2 structure) to 0.25 eV (AA1 structure), shown in Table I, to be compared with 0.64 eV for monolayer $C_3B$. These results suggest that interlayer chemical interactions have strong effects on the band edge states, resulting in a band gap reduction of as much as 0.56 eV for the bilayer system at the DFT-PBE level. Similar effects are also observed in a recent work on $C_3B$/$C_3N$ bilayer system [44]. After including the GW quasiparticle corrections, the band gaps increase to 1.66 to 1.89 eV as shown in Table II. The presumably weak interlayer interaction thus induces a reduction of quasiparticle band gap of about 0.9 eV in the bilayer system. The surprisingly large band gap variation (0.9 eV) from monolayer to bilayer system is largely due to the fact that the band edge states have significant $p_z$ character which is prone to out-of-plane perturbations.

TABLE II. DFT-PBE and GW band gaps (in eV) of monolayer, bilayer, and bulk $C_3B$. $\Delta E_g$ is the quasiparticle corrections to the band gap.

|  | Monolayer | Bilayer | | | | Bulk | | | |
| --- | --- | --- | --- | --- | --- | --- | --- | --- | --- |
|  |  | AA1 | AA2 | AB1 | AB2 | AA1 | AA2 | AB1 | AB2 |
| PBE | 0.64 | 0.25 | 0.10 | 0.24 | 0.08 | -0.38 | -0.68 | -0.19 | -0.47 |
| GW | 2.55 | 1.86 | 1.66 | 1.89 | 1.66 | 0.62 | 0.17 | -- | -- |
| $\Delta E_g$ | 1.91 | 1.61 | 1.56 | 1.65 | 1.58 | 1.00 | 0.85 | -- | -- |

We proceed to investigate bulk $C_3B$ using the same stacking models as those for the bilayer system shown in Figure 5. Table III shows the optimized structural parameters and relative energies for these models. Similar to the bilayer systems, the AA2 stacking is predicted to be the lowest energy structure which is again in sharp contrast to the results for $C_3N$ [16] for which the AA2, AB1, and AB2 structures are practically degenerate. The interlayer separation decreases slightly in the bulk phases compared with the respective bilayer systems. In their original work [18], Kouvetakis *et al.* did not observe a dominant layer stacking pattern for the as-synthesized bulk $C_3B$ without additional post-synthesis treatments. Considering the substantial energetic difference between different stacking orders, it might be possible to obtain perfectly AA2 stacked single crystal $C_3B$ with the assistance of a post-annealing process.

TABLE III. Crystal structures (*a* is the in-plane lattice constant and *d* is the interlayer distance) and their relative energies (meV/atom) for the four bulk models.

| Functional | PBE | | | vdW-optB86b | | | SCAN+rVV10 | | |
| --- | --- | --- | --- | --- | --- | --- | --- | --- | --- |
|  | *a* | *d* | $\Delta E$ | *a* | *d* | $\Delta E$ | *a* | *d* | $\Delta E$ |
| Stacking | (Å) | (Å) | (meV) | (Å) | (Å) | (meV) | (Å) | (Å) | (meV) |
| AA1 | 5.166 | 3.821 | 9.65 | 5.162 | 3.502 | 29.28 | 5.134 | 3.468 | 26.27 |
| AA2 | 5.168 | 3.608 | 0.00 | 5.166 | 3.248 | 0.00 | 5.135 | 3.201 | 0.00 |
| AB1 | 5.165 | 3.705 | 18.73 | 5.163 | 3.342 | 29.65 | 5.135 | 3.350 | 26.30 |
| AB2 | 5.165 | 3.591 | 14.01 | 5.167 | 3.163 | 14.57 | 5.138 | 3.178 | 12.68 |

Figure 7 shows the DFT-PBE band structures for the four bulk stacking models. All structures show semi-metallic band structure behavior with small overlaps between valence and conduction bands. Note that in Table II these overlaps are shown as negative band gaps to emphasize the semi-metallic nature of the DFT band structures. These results agree with previous theoretical work [20] which also seem to be consistent with experiment [18]. However, considering that DFT-PBE almost always underestimates the band gap of *sp* semiconductors, it is likely that ideal bulk $C_3B$ with a perfect stacking order be a narrow-gap semiconductor. To this end, we have carried out GW calculations for the AA1 and AA2 models since both have a small unit cell of 8 atoms. In addition, the AA2 stacking also has the lowest energy among the four models.

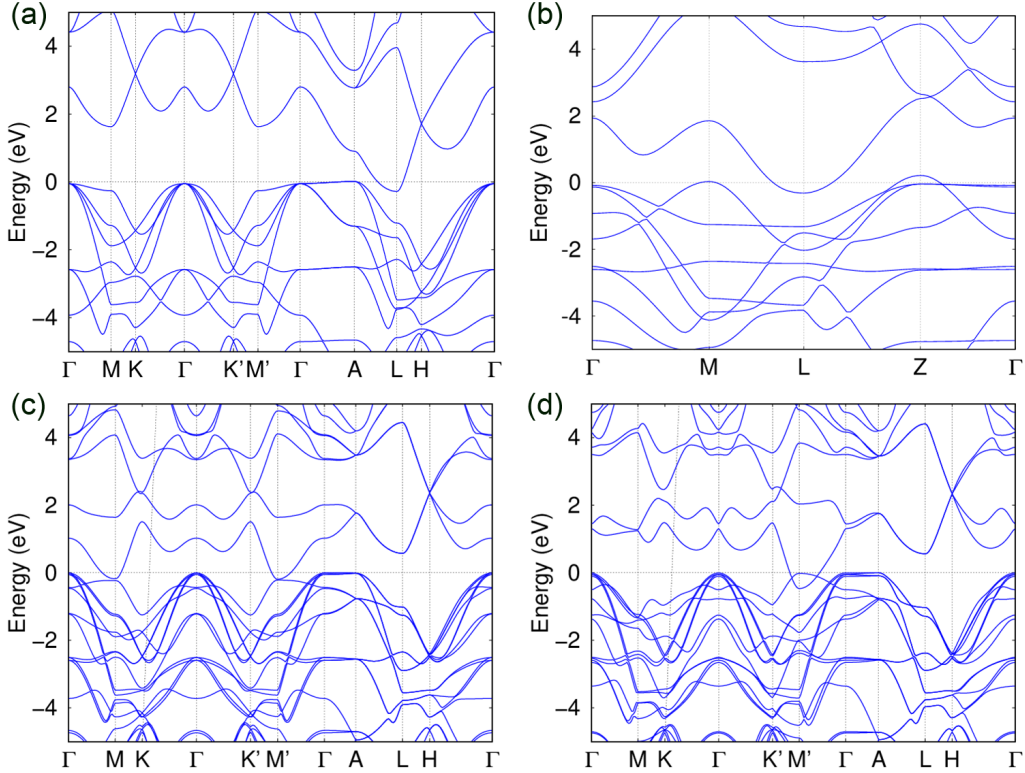

FIG. 7. DFT-PBE band structures of bulk $C_3B$: (a) AA1, (b) AA2, (c) AB1, and (d) AB2.

Figure 8 shows the GW quasiparticle band structures for the AA1 and AA2 models. Not surprisingly, both structures are predicted to be narrow-gap semiconductors. In particular, the GW band gap of the lowest-energy AA2 structure is only 0.17 eV as shown in Table II. The quasiparticle correction to the band gap is, however, significantly reduced compared with that of monolayer (0.85 vs 1.91 eV) due to the enhanced dielectric screening in the bulk phase. Thus, interlayer coupling and reduced quasiparticle corrections both contributed to the dramatic drop (about 2.38 eV) in band gap going from monolayer to bulk structures. To the best of our knowledge, such a large band gap renormalization has not been observed in other van der Waals bonded layered materials. We would also like to mention that it is not surprising that the as-synthesized bulk $C_3B$ appeared to be (semi)metallic [18] considering that the sample may contain substantial amount of defects and impurities, which, when coupled with stacking disordering and temperature effects, may give rise to the metallic appearance of the samples. In addition, there is a limitation of the accuracy of current GW approach, which is likely about 0.1 eV, making a definite prediction of narrow-gap semiconductors difficult. Our results call for future experimental verifications. Note that we do not carry out GW calculations for the AB1 and AB2 models due to the entanglement of the conduction and valence bands which make subsequent GW calculations more difficult. On the other hand, we feel that it is not necessary to do calculations for all bulk structures since the GW correction to the band gap would likely be similar for AA1 and AA2 structures.

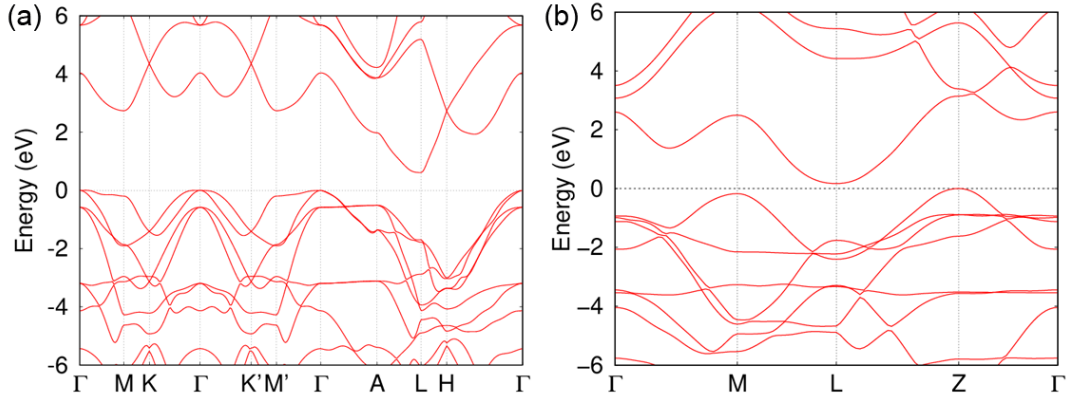

FIG. 8. GW band structures of bulk $C_3B$ with the AA1 (a) and AA2 (b) stacking orders. Both systems are predicted to be narrow-gap semiconductors.

## C. Tuning the valence and conduction band edge states separately through in-plane and out-of-plane strains.

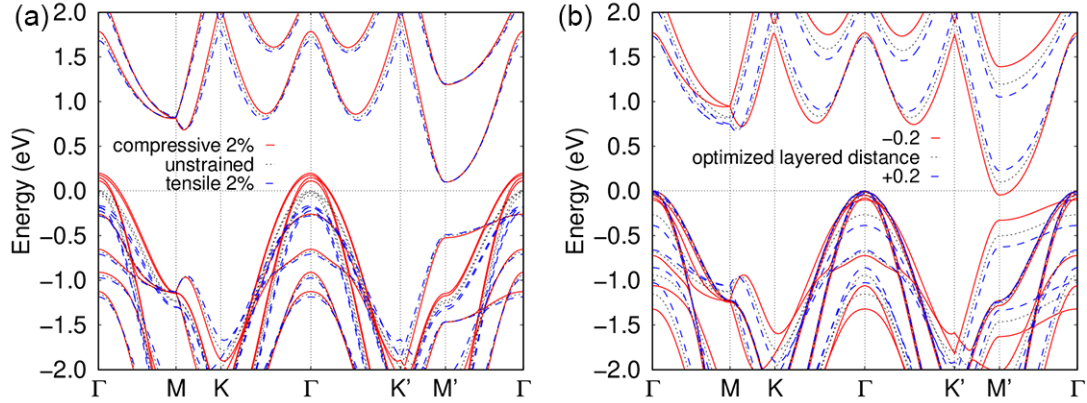

FIG. 9. DFT band structures of the AA2 bilayer structure calculated with (a) in-plane strains and (b) artificially modulating the interlayer distance, illustrating the contrasting and distinct tunability of the band-edge states.

To further demonstrate the contrasting and distinct tunability of the band-edge states, we have carried out PBE band structure calculations for the AA2 bilayer structure (the most stable bilayer phase as shown in Figure 5 and Table I) under in-plane and out-of-plane strains. Figure 9 (a) compares band structures of the AA2 bilayer structure calculated using optimized (unstrained) structure and structures with ±2% bi-axial in-plane strains. Whereas out-of-plane $p_z$ derived states are mostly unaffected by the in-plane strain, the $p_{xy}$ derived VBM states are shifted by ±0.18 eV with the ±2% bi-axial in-plane strains, suggesting a large valence band deformation potential $\delta\varepsilon/(\delta a/a) = 9.0$ eV. In contrast, the $p_z$ derived states are strongly affected by the interlayer separation as shown in Figure 9 (b), in particular, the bonding and anti-bonding $p\pi$ states which form minimum direct gap at the M' point. The minimum direct gap at the M' point can be tuned by ±0.35 eV with a ±0.2 Å change to the interlayer distance.

It should be mentioned that proper alignment of the band structures of different systems (in this case, structures with different strains) is a challenging problem. Here we do not intend to quantify the absolute changes to the band edge states under strain. Instead, we would like to illustrate how band edge states with contrasting wave function characteristics (i.e., in-plane vs. out-of-plane) respond differently to strains, and how this property can be used to tune band edge states using in-plane and out-of-plane strains. When comparing the band structures calculated with different in-plane strains, we first place the VBM of the unstrained structure at zero. We then align the CBM levels of the band structures calculated with different in-plane strains. The reason for aligning CBM levels is that the CBM is derived from $p_z$ orbitals as we have shown in Fig. 3; these states (as well as $p_z$ orbitals derived valence states) are insensitive to in-plain strains. Therefore, by aligning $p_z$ orbitals derived states, the effects of strain on the $p_{x,y}$ states can be clearly seen. For out-of-plane strains, the band structures are aligned at the VBM. The out-of-plain strains mostly affects the $p_z$ orbitals derived states, as it is clearly illustrated in the right panel of Fig. 9. The GW quasiparticle corrections are not expected to change significantly with strains.

### D. Convergence behavior of GW calculations for 2D materials

We now discuss the convergence behavior of GW calculations for 2D materials. The challenge of carrying out fully converged GW calculations for 2D materials is well-recognized and has been discussed extensively. First, one needs to ensure that the GW results are converged with respect to the number of bands included in the GW calculations as well as the kinetic energy cutoff of the dielectric matrices. Figure 10 (a) shows the convergence behavior of the calculations minimum (indirect) gap of $C_3B$ with respect to these truncation parameters. A kinetic cutoff of about 30 Ry for the dielectric matrix is needed to converge the calculated band gap to within 0.02 eV. In addition, one needs to include about 10,000 bands (lower horizontal values) in the calculations of the dielectric matrix and self-energy using the conventional band summation method. Using the energy-integration method [30], the number of integration points (bands) is reduced to about 500 as shown in Figure 10 (a) (upper horizontal values). Therefore, we are able to reduce the computational cost by well over one order of magnitude for these systems.

Another difficulty comes from the Brillouin zone (BZ) integration of the GW self-energy, which is often carried out on a uniform sampling grid: $\Sigma(n\vec{k}) = \sum_{\vec{q}} f_{\vec{q}} \Sigma_{n\vec{k}}(\vec{q})$, where $\Sigma(n\vec{k})$ is the GW self-energy for state $n\vec{k}$. This summation usually converges rather quickly with respect to the BZ sampling density for bulk (3D) semiconductors. For 2D materials, however, the convergence is extremely slow [32-35,45,46] due to the analytical behaviors of the 2D dielectric function as the wave vector $\vec{q}$ approaches 0. The red curve in Figure 10 (b) shows the calculated band gap of $C_3B$ as a function of the 2D BZ sampling density using the conventional uniform sampling technique. One needs at least an $18 \times 18 \times 1$ $k$-grid to properly converge the band gap for this system.

Recently, we have developed a new technique [16,29] that can significantly reduce the required BZ sampling density in 2D GW calculations. Our method is inspired by the recent work [46] in which the authors proposed a non-uniform subsampling technique to combat the slow convergence problem in 2D GW calculations. In our new method, the long wave length (i.e., small $q$) contribution to the self-energy of state $n\vec{k}$ is replaced with an analytical integration over the mini-BZ enclosing the $\Gamma$ (i.e., $\vec{q}=0$) point. Briefly, we first calculate the self-energy $\Sigma_{n\vec{k}}(\vec{q})$ for a few small $q$ points (typically 3 or 4 points). The results are then fitted with some analytical function; the integration over the mini-BZ can then be calculated analytically: $\Sigma(n\vec{k}) = \sum_{\vec{q}\neq 0} f_{\vec{q}} \Sigma_{n\vec{k}}(\vec{q}) + \frac{f_0}{A_{mBZ}} \int_{mBZ} \Sigma_{n\vec{k}}(\vec{q}) d^2\vec{q}$, where $A_{mBZ}$ is the area of the mini-BZ. Using this method, we are able to achieve well converged GW results using a $6\times 6\times 1$ k-grid as shown with the blue curve in Figure 10 (b). Note that the computational cost for calculating the dielectric function scales as $O(N_k^2)$, where $N_k$ is the number the BZ integrating points. Our method represents a speedup factor of nearly two orders of magnitude $[(18^2/6^2)^2 = 81]$.

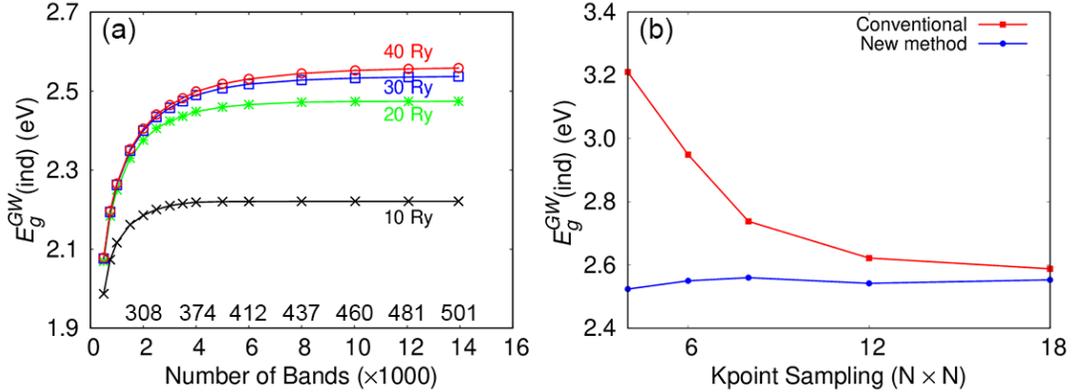

FIG. 10. Convergence behavior of the calculated GW band gap for monolayer $C_3B$. Panel (a) shows the convergence with respect to the number of bands included in the self-energy summation and the cutoff for the dielectric matrix; panel (b) compares the BZ integration convergence behavior using the conventional (shown in red) and the new integration approach (blue). Note that there are two sets of numbers on the horizontal axis in panel (a) as explained in the main text.

We would like to mention that although different plasmon-pole models (or without the use of a plasmon-pole model) for the dielectric function in GW calculations may yield somewhat different convergence behavior [47], the convergence issues that we discussed in our work are always relevant regardless of the particular model/approach used. In other words, even if other plasmon-pole models are used, one should still carefully check the relevant convergence parameters that we discuss here. Note that the mini-BZ sub-sampling and analytical integration approach we have developed to accelerate the BZ integration for 2D GW calculations should only be used for 2D semiconductors since metallic systems would still require a fairly dense $k$-grid to accurately capture the intra-band transitions. In addition, the dielectric function and electron self-energy of 2D metallic systems have different asymptotic behavior in the small $q$ limit.

## IV. Conclusion

Practical applications of 2D materials in fields such as flexible electronics, nano-photonics, and nano-sensing and actuating require a sizable band gap. In fact, difficulties in obtaining a stable and controllable band gap in graphene has greatly hindered the application of this otherwise promising material. A single 2D semiconductor with a conveniently tunable band gap is of particular interest since it may be easily adapted to operate at different electrical voltage or optical wavelength. Graphitic boron carbide $C_3B$ is an emerging layered material that has been successfully synthesized in the bulk form [18] and has recently been predicted [19] to be a stable 2D semiconductor. Indeed, there is a large family of graphitic boron carbide ($C_xB$) and carbon nitride ($C_xN$) semiconductors that awaits discovery. Unfortunately, accurate understanding of the layer-dependent electronic properties of $C_3B$ and other graphitic boron carbide semiconductors is still lacking.

In this work, we have carried out detailed DFT+GW studies of the electronic and structural properties of $C_3B$. Our fully converged GW calculations (with respect to the number of bands, energy cutoff and $k$-point sampling) predict that monolayer $C_3B$ is a semiconductor with a moderate quasiparticle band gap of about 2.55 eV, whereas that of a perfectly stacked bulk phase is as small as 0.17 eV, giving rise to a band gap renormalization of over 2.3 eV due to the interlayer interaction and screening effects. Thus, our results suggest that $C_3B$ could be an interesting layered semiconductor with a remarkably band gap tunability through dimensionality. To the best of our knowledge, such a large band gap renormalization has not been observed and/or predicted in other van der Waals bonded layered materials. The calculated quasiparticle band gap of monolayer $C_3B$ is over 1.0 eV larger than that of $C_3N$, a closely related 2D semiconductor. Detailed inspections of the atomic characters of the near-edge electronic states reveal that the band gap is formed between out-of-plane CBM and in-plane VBM states. This raises an interesting possibility of tuning the VBM and CBM states separately by in-plane and out-of-plane interactions. The fact that the properties of valence and conduction states may be individually controlled offers another interesting knob to tailor this material for practical applications. For example, in 2D heterostructures, it is sometimes desirable to tune the valence or conduction band energy to realize different band offset schemes.

**Acknowledgement**

This work is supported in part by the National Natural Science Foundation of China (Nos. 51632005, 51572167, and 11929401), and the National Key Research and Development Program of China (No. 2017YFB0701600). Work at UB is supported by the US-NSF under Grant No. DMR-1506669 and No. DMREF-1626967. P. Z. acknowledges the Southern University of Science and Technology (SUSTech) of China for supporting his extended visit during which part of this work was done. W. Z. also acknowledges the support from the Guangdong Innovation Research Team Project (Grant No. 2017ZT07C062), and the Shenzhen Pengcheng-Scholarship Program. We acknowledge the computational support provided by the Center for Computational Science and Engineering at Southern University of Science and Technology, the Beijing Computational Science Research Center, and the Center for Computational Research at UB.